# Variable Electron-Phonon Coupling in Isolated Metallic Carbon Nanotubes Observed by Raman Scattering

Yang Wu[1], Janina Maultzsch[1], Ernst Knoesel[2], Bhupesh Chandra[3], Mingyuan Huang[3], Matthew Y. Sfeir[4], Louis E. Brus[4], J. Hone[3], and Tony F. Heinz[1]

[1]*Depts. of Physics and Electrical Engineering, Columbia University, New York, NY 10027*

[2]*Dept. of Physics, Rowan University, Glassboro, NJ 08028*

[3]*Dept. of Mechanical Engineering, Columbia University, New York, NY 10027*

[4]*Dept. of Chemistry, Columbia University, New York, NY 10027*

**Abstract:**

We report the existence of broad and weakly asymmetric features in the high-energy (*G*) Raman modes of freely suspended metallic carbon nanotubes of defined chiral index. A significant variation in peak width (from 12 cm$^{-1}$ to 110 cm$^{-1}$) is observed as a function of the nanotube's chiral structure. When the nanotubes are electrostatically gated, the peak widths decrease. The broadness of the Raman features is understood as the consequence of coupling of the phonon to electron-hole pairs, the strength of which varies with the nanotube chiral index and the position of the Fermi energy.

Raman scattering from high-energy optical phonons in metallic single-walled carbon nanotubes (SWNTs) has been the subject of much recent interest [1-12]. The Raman spectra of these high-energy modes (HEMs) or *G* modes typically display a distinctive metallic line shape. In contrast to the sharp, symmetrical lines for these modes in semiconducting SWNTs, metallic nanotubes in bundles have generally been reported to exhibit broadened and asymmetric Raman line shapes [1-3]. These metallic line shapes provide an important signature of the fundamental characteristics of phonons in metallic nanotubes. Understanding the corresponding coupling between optical phonons and low-energy electronic excitations is also of interest because of its role in charge transport, particularly under conditions of high bias voltage or current saturation [13-16]. Further, from the practical side, observation of broadened HEM Raman features has often been used as a spectroscopic method for the identification of metallic SWNTs [17].

Despite this motivation to develop a full understanding of Raman scattering from the HEMs in metallic nanotubes, several basic features of the process have remained unclear. Indeed, there have been contradictory experimental reports in the literature about whether the line broadening is an intrinsic feature of metallic SWNTs [5, 7, 8] or arises only as a consequence of bundling, surface interactions, or other perturbations [6, 9]. The same controversy exists among theoretical descriptions of the phenomenon, which attribute the broadening either to electron-phonon coupling [12] or to plasmons in nanotube bundles [10]. Similarly, while asymmetric lines of a Breit-Wigner-Fano (BWF) form have frequently been reported [2, 3], other researchers did not observe such asymmetry [8]. To resolve these uncertainties, we have probed isolated metallic SWNTs that are suspended in air and are thus free of environmental perturbations from the substrate and other nanotubes in bundles. Making use of Rayleigh scattering in conjunction with Raman measurements of the radial breathing mode (RBM), we determined the chiral indices ($n,m$) of specific nanotubes. This permitted us to analyze HEM line shape with independent knowledge of the nanotube's metallic or semiconducting character. In addition, we applied electrostatic gating to the metallic nanotubes to observe the influence of modifying their low-lying electronic transitions.

In this Letter, we demonstrate directly the existence of broad and weakly asymmetric (BWF) line shapes in the HEM of isolated metallic nanotubes. The line broadening and asymmetry are thus *intrinsic* properties of metallic SWNTs. The data also reveal that while line broadening is typically present in the Raman spectra of the HEM, there is actually a significant dependence of the line shape on the chiral structure of the SWNT. This observation may explain the seemingly contradictory results that have appeared in the literature. The broadening of the HEM Raman feature is sharply decreased by shifting the Fermi energy of the nanotube through electrostatic gating. This finding shows the critical role of the resonant low-energy electronic excitations. We interpret the broadening of HEM Raman features in metallic nanotubes as arising from the strong coupling of optical phonons to resonant excitation of electron-hole pairs, an effect that has been predicted to be significant theoretically [12]. If we attribute the width of the HEM Raman features entirely to such electronic damping, we infer lifetimes of zone-center optical phonons as short as 100 – 200 fs.

The samples were isolated SWNTs suspended over a 100-$\mu$m open slit in air. These nanotubes were prepared by chemical vapor deposition on substrates with an etched slit structure [18]. Typical diameters of the nanotubes in our study were in the range of 1.9 – 2.4 nm. In addition to providing us with isolated, unperturbed SWNTs in which to probe the Raman response, the suspended SWNT samples permitted optical characterization by Rayleigh scattering [19, 20]. For each nanotube we selected a Raman excitation wavelength that was resonantly enhanced by an electronic transition. We used laser lines at wavelengths of 514 nm, 532 nm, and 632 nm. Determination of the nanotube chiral indices relied on both the RBM frequency [21] and the electronic spectra from Rayleigh scattering. These assignments are supported by earlier investigations in which Rayleigh spectra were measured for nanotubes with structure independently determined by transmission electron microscopy [20]. In addition, the assignments are consistent with Kataura plot patterns found by photoluminescence and resonant Raman spectroscopy [22-26]. Although the assignments do not rely on absolute values of the transition energies predicted by theory, they agree with empirically adjusted tight-binding models that show those trends and patterns [27, 28]. To examine the effect of a shift in the Fermi energy, we held a fine gating electrode at a distance of 20-40 $\mu$m from the suspended nanotube. We performed Raman measurements while a potential was applied between the gate and the nanotube (through Cr/Au contacts on the substrate).

Figure 1 shows spectroscopic data for five metallic nanotubes with the indicated chiral indices, arranged in order of increasing chiral angle. The left column displays the Rayleigh scattering spectra; the right column presents the Raman spectra of the HEM, with insets showing RBM spectra. The features in the Rayleigh spectra arise from the second metallic transitions ($M_{22}$). Except for the case of armchair nanotubes [$(n,n)$], trigonal warping causes this transition to be split into lower- ($M_{22}^-$) and higher- ($M_{22}^+$) energy levels [29]. This double-peaked structure is seen in all of the Rayleigh spectra of Fig. 1, except for the top and bottom panels. The bottom spectrum is assigned to an armchair nanotube for which no splitting is expected, while the $M_{22}^+$ peak of the nanotube in the upper panel falls outside the spectral range of the measurement.

We now turn to a discussion of the line shape of the HEM Raman features. The spectra have, in general, two distinct components. There is a lower-frequency peak (P1) that is broad and often exhibits asymmetry. This peak, with a line shape similar to those in the upper three panels, is usually referred to as "metallic" in character and shows variation with the nanotube chiral indices (*n,m*). A higher-frequency peak (P2) is also generally present; it is narrow and symmetrical. We model the contribution of P1 by the BFW form and P2 by a Lorentzian. The spectra for the HEMs at a Raman shift of frequency $\omega$ are thus described by

$$I(\omega) = I_1 \frac{\left[1 + 2(\omega - \omega_1)/q\Gamma_1\right]^2}{1 + \left[2(\omega - \omega_1)/\Gamma_1\right]^2} + I_2 \frac{1}{1 + \left[2(\omega - \omega_2)/\Gamma_2\right]^2} \quad (1)$$

Here $\Gamma_1$, $\Gamma_2$; $I_1$, $I_2$; and $\omega_1$, $\omega_2$ are, respectively, the full widths at half maximum (FWHM), the line strengths, and the phonon frequencies for the P1 and P2 components; $q$ is the asymmetry parameter for the BWF form of P1. As can be seen from the fits of Fig. 1, this form provides a good description of our experimental data. Table I summarizes the fitting parameters, as well information used for the index assignments.

From these data on individual, suspended nanotubes, we can immediately make the following observations: (*i*) The broadening of the HEM is an *intrinsic* property of metallic carbon nanotubes and is not induced by the environment or nanotube-nanotube interactions. For these isolated, suspended nanotubes we see a width $\Gamma_1$ extending to > 100 cm$^{-1}$. The typical width for these modes in semiconducting nanotubes (and the average width for peak P2), on the other hand, is ~ 10 cm$^{-1}$ [5]. (*ii*) The asymmetry of the line shape is also intrinsic in nature. For most of the nanotubes we investigated, we observed a weak but clear asymmetry, as indicated by the asymmetry parameters *q* in Table I. A detailed analysis of this asymmetry and its dependence on excitation energy will be presented elsewhere. (*iii*) Although intrinsic, the broadening of the Raman HEM of metallic nanotubes exhibits strong variation with the (*n,m*) species, with overall widths ranging from 110 cm$^{-1}$ [(24,0) tube] to 12 cm$^{-1}$ [(15,15) tube].

Observation (*iii*) has two immediate implications. First, the wide variation in the degree of the broadening of the HEMs for metallic nanotubes offers an explanation for some of the apparently contradictory assertions in the literature. Figure 1 shows that metallic nanotubes exhibit significant or moderate or negligible intrinsic broadening, depending on the nanotube chiral index. Our observed peak widths fall into the range of previously reported widths of ~10-100 cm$^{-1}$ from Raman experiments on individual metallic nanotubes [4, 5, 7-9], although our nanotube diameters are somewhat larger than those in most studies. The second conclusion is related to the use of the broadening of the HEM Raman lines as a way to identify metallic nanotubes. Our measurements show that broadening is a *sufficient*, but *not a necessary* signature of a metallic nanotube. The one-to-one correspondence between line broadening and metallic character is thus invalid.

The two observed features of the HEM are generally associated with the longitudinal optical (LO) and transverse optical (TO) modes. In chiral nanotubes, both of them are Raman active; their character is not purely longitudinal or transverse [30]. For the two high-symmetry structures of zigzag [(*n,0*)] and armchair [(*n,n*)] nanotubes, however, special selection rules apply [31]. For the former, only the LO mode is Raman active; while for the latter, only the TO mode is allowed [21]. We consequently assign the broad feature in the zigzag nanotube to the LO mode and the narrow peak in the armchair nanotube to the TO mode. By extension, we associate the broad P1 and the narrow P2 with the LO and TO modes, respectively, as predicted from first-principles calculations [12, 32]. This interpretation of the HEM also agrees with the discussion in [7] for the case of resonant excitation. The unexplained variation of the broadened peak P1 with (*n,m*) could be related to two effects: One is the variation of the electronic structure of metallic nanotubes with chiral angle. In particular, a small gap opens in the electronic structure of non-armchair metallic nanotubes. Another effect might be the variation of the phonon displacements with changing nanotube chirality [30] and its potential influence on the electron-phonon coupling.

Having now established the intrinsic character of broadening of the HEM in metallic nanotubes, how do we understand the origin of this effect? From the absence of similar line broadening in semiconducting nanotubes, we can immediately infer that the low-energy electronic excitations of the metallic nanotubes play the decisive role. We therefore interpret the broadening as arising from coupling between the phonon and the electronic excitations, in accordance with theoretical predictions [12]. Note that this electron-phonon coupling is not related to the coupling of the Raman-active phonons to optically excited electrons, which determines the intensity of the Raman signal.

In any metallic system there are low-energy excitations that can match the energy of a phonon. In (first-order) Raman scattering, however, we observe only zone-center phonons, which cannot normally undergo direct decay into electron-hole pairs: Low-energy electronic transitions in a normal metallic band are available only with significant momentum change. For the case of metallic nanotubes, with linear bands that are occupied up to their crossing (Dirac) point, we do have available vertical electronic transitions at energies of optical phonons. Given the magnitude of the pseudo-gap for chiral metallic nanotubes, the same situation applies when the bands are not strictly linear.

To test the above scenario experimentally, we change the occupancy of the metallic bands. If the Fermi energy is moved either up or down by half the phonon energy, the relevant vertical transitions will, neglecting the effect of finite temperature, be blocked. This is shown schematically in Fig. 2 (a). In Fig. 2 (b) we plot the peak width $\Gamma_1$ of the HEM of a (22,10) SWNT as a function of gating voltage. The width $\Gamma_1$ decreases for sufficiently large positive or negative gating voltages, as expected from a displacement of the Fermi energy. In addition, the frequency of P1 exhibits a slight blue shift when the width decreases. A similar effect has been observed recently in graphene [33, 34].

The experimental results for the dependence of the HEM linewidth on the gating voltage can be reproduced with the following model. Considering the Raman broadening to be proportional to the statistical availability of electron-hole pair generation at the phonon energy, we have

$$\Gamma(E_F) = \Gamma_0 + \Gamma_{e-ph} f(T, E_F, -\hbar\omega/2)[1 - f(T, E_F, \hbar\omega/2)] \qquad (2)$$

Here $\Gamma_{e-ph}$ denotes the maximum broadening from the electron-phonon coupling, $\Gamma_0$ is the contribution from phonon-phonon coupling and other sources independent of electronic effects; Fermi factors $f$ are expressed as a function of temperature $T$, Fermi energy $E_F$ and phonon energy $\hbar\omega$. We assume that $E_F$ scales linearly with the gating voltage, with a possible constant offset from initial doping of the nanotube in air. The predicted linewidth as a function of gate bias (gray curve in Fig. 2) matches the experimental data reasonably well with $T = 115^{o}C$. The gating dependence of the HEM Raman scattering highlights the importance of the particular band structure of metallic nanotubes: the low-energy vertical electronic transitions, which are not a generic feature of metals, can be turned off by moving the Fermi energy away from the Dirac point. Furthermore, our results show that the essential feature of a "metallic" HEM is its width (given by the electron-phonon coupling), as opposed to the asymmetry which is less apparent and might be even absent in spite of strong electron-phonon coupling.

In conclusion, we have demonstrated the existence of intrinsically broad (and generally asymmetric) line shapes for Raman scattering of the high-energy (G) modes in isolated metallic carbon nanotubes. The broadening originates from the coupling between the relevant zone-center phonons and vertical (zero-momentum) electronic transitions that are available in metallic nanotubes. The strength of the coupling, as manifested by the Raman line width, can be modified by shifting the nanotube Fermi energy through electrostatic gating. A significant variation in the degree of coupling is also observed as a function of the chiral structure of the metallic nanotube. Therefore broadened Raman lines are a sufficient but not a necessary signature to identify a metallic nanotube.

We acknowledge support from the Nanoscale Science and Engineering Initiative of the NSF under Award Nos. CHE-0117752 and ECS-05-07111, the New York State Office of Science, Technology, and Academic Research (NYSTAR), and the Office of Basic Energy Sciences, U.S. DOE (grants DE-FG02-98ER-14861 and DE-FG02-03ER15463), Intel Corporation, and from the Alexander-von-Humboldt Foundation for JM.

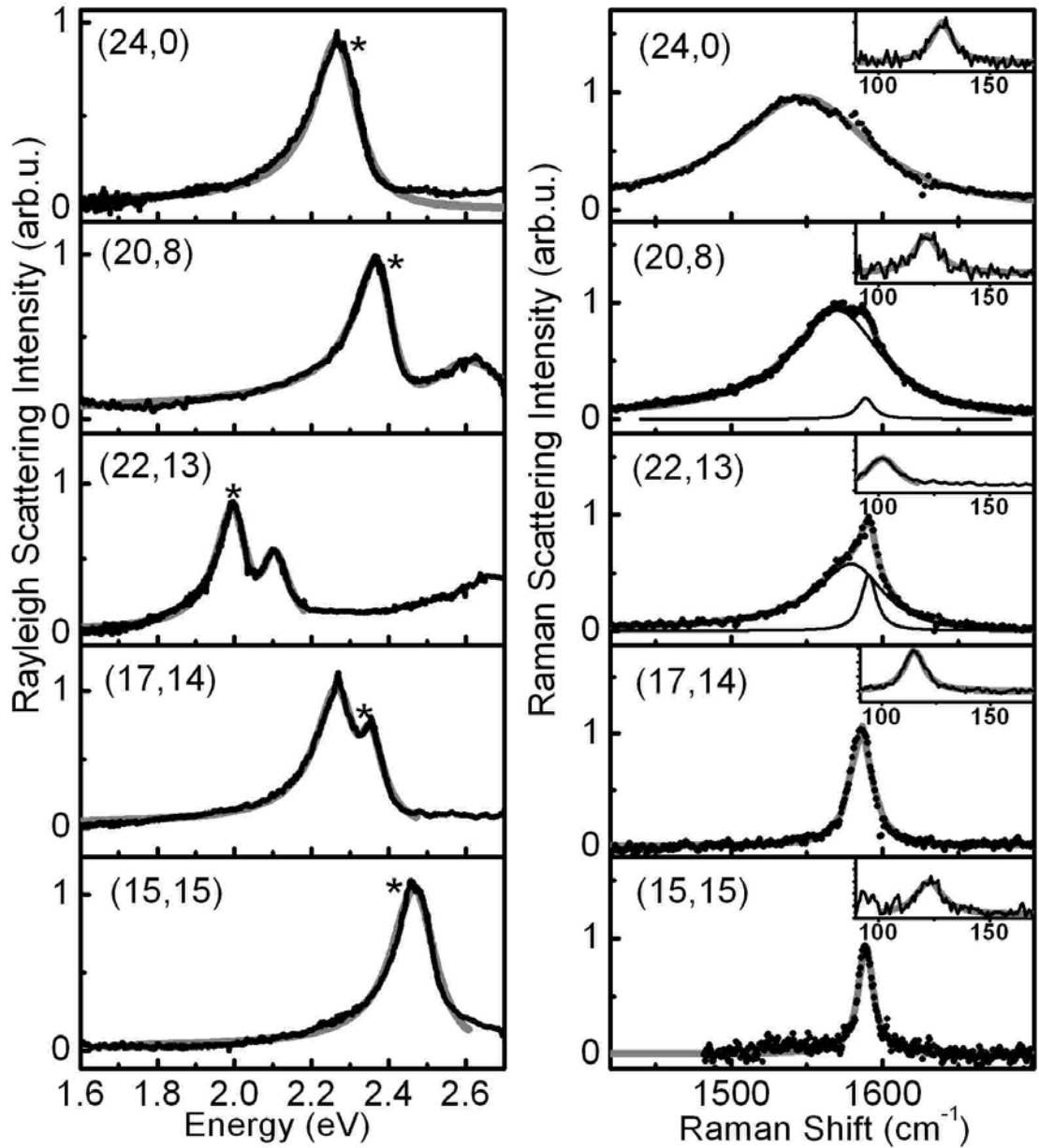

**Fig. 1:** Rayleigh (left column) and Raman (right column) spectra of the HEM for five individual metallic nanotubes of the indicated chiral structure. The gray curves for the Rayleigh spectra are based on Lorentzian absorption profiles [19]. The stars in the Rayleigh spectra indicate the laser energy for the Raman measurements. The gray lines in the right panel are fits to Eq. (1), with parameters as indicated in Table I. Insets: RBM Raman spectra.

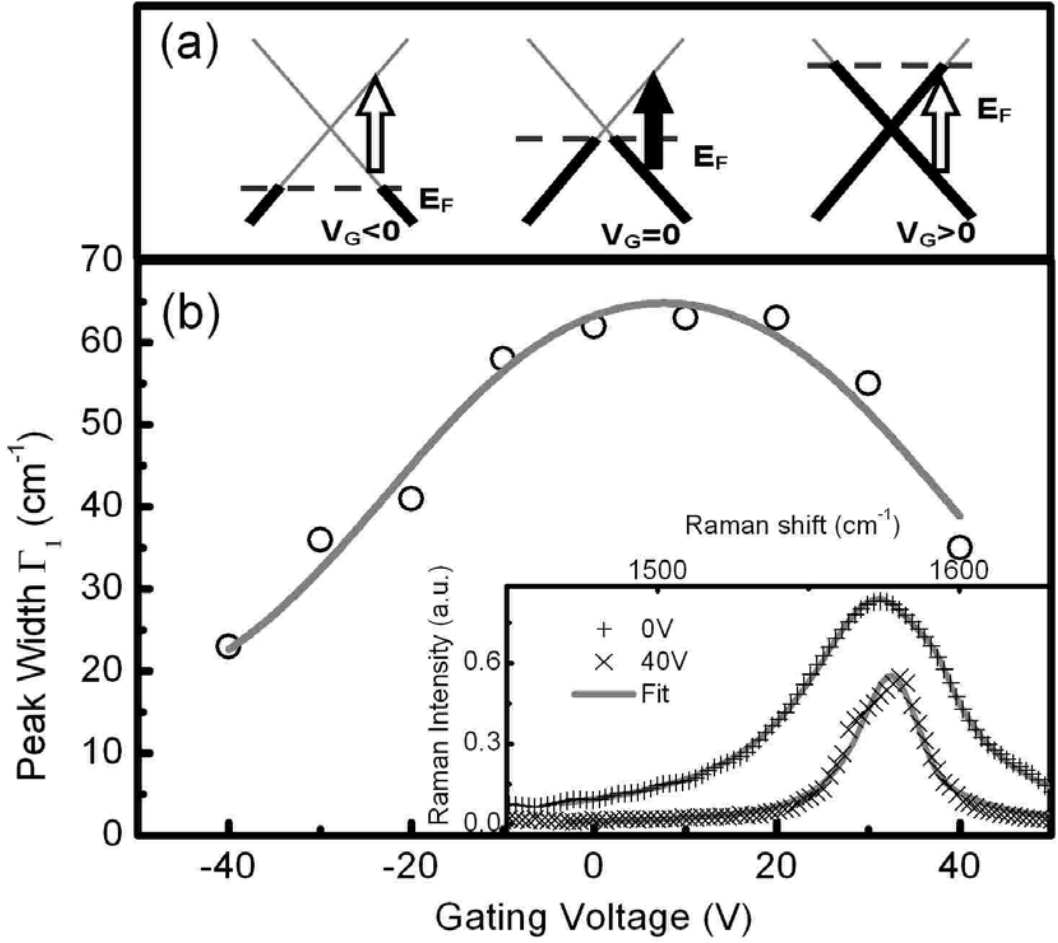

**Fig. 2:** Raman scattering of the HEM of a metallic nanotube under electrostatic gating. (a) Schematic illustration of blocking of vertical low-energy electronic transitions when the nanotube is sufficiently biased at either positive or negative potential. (b) Full width $\Gamma_1$ of peak P1 from a (22,10) nanotube as a function of gating voltage. The gray line is a fit based on the model discussed in the text with $\hbar\omega = 1588$ cm$^{-1}$, T = 115$^{\circ}$C, $\Gamma_0 = 12$ cm$^{-1}$, and $\Gamma_{el\text{-}ph} = 62$ cm$^{-1}$. Inset: Raman spectra for two different gating voltages, as measured with 2.33 eV photons.

| Nanotube Assignment | | | Exp. Rayleigh | | Theory | | RBM Raman $(cm^{-1})$ | HEM Raman | | |
|---|---|---|---|---|---|---|---|---|---|---|
| $(n,m)$ | $\theta(°)$ | $d$ (nm) | $M_{22}^{(-)}$ (eV) | $M_{22}^{(+)}$ (eV) | $M_{22}^{(-)}$ (eV) | $M_{22}^{(+)}$ (eV) | | $\Gamma_1$ $(cm^{-1})$ | $\Gamma_2$ $(cm^{-1})$ | $q$ |
| (24,0) | 0 | 1.91 | 2.30 | -- | 2.36 | 2.80 | 129 | 110 | -- | -15 |
| (20,8) | 16.1 | 1.98 | 2.36 | 2.62 | 2.34 | 2.59 | 122 | 74 | 12 | -16 |
| (22,10) | 17.8 | 2.25 | 2.12 | 2.30 | 2.13 | 2.30 | 111 | 56 | 12 | -23 |
| (22,13) | 21.6 | 2.43 | 2.00 | 2.10 | 2.02 | 2.13 | 101 | 48 | 12 | -10 |
| (17,14) | 26.8 | 2.13 | 2.28 | 2.35 | 2.28 | 2.33 | 115 | -- | 18 | -- |
| (15,15) | 30.0 | 2.06 | 2.45 | 2.45 | 2.37 | 2.37 | 122 | -- | 12 | -- |

**Table I:** Summary of data for the nanotubes presented in Figs. 1 and 2. The table shows the assigned nanotube chiral index and the corresponding chiral angle and diameter; the electronic transition energies for the $M_{22}$ feature from the Rayleigh scattering data and from a non-orthogonal tight-binding model [28] with the previously established empirical adjustment of all energies upwards by 300 meV [27]; the measured RMB frequency; and the fitting parameters for the measured HEM Raman spectra, with the full-widths of the two peaks and the asymmetry parameter of peak P1 as defined by Eq. (1). The (24,0) zigzag and (15,15) armchair nanotubes, as well as the (17,14) near-armchair nanotube, are described by a single peak.

**References**

[1] M. A. Pimenta *et al.*, Phys. Rev. B 58, 16016 (1998).
[2] Z. H. Yu, and L. Brus, J. Phys. Chem. B 105, 1123 (2001).
[3] S. D. M. Brown *et al.*, Phys. Rev. B 6315, 155414 (2001).
[4] A. Jorio *et al.*, Phys. Rev. B 65, 155412 (2002).
[5] A. Jorio *et al.*, Phys. Rev. B 66, 115411 (2002).
[6] C. Jiang *et al.*, Phys. Rev. B 66, 161404 (2002).
[7] J. Maultzsch *et al.*, Phys. Rev. Lett. 91, 087402 (2003).
[8] M. Oron-Carl *et al.*, Nano Lett. 5, 1761 (2005).
[9] M. Paillet *et al.*, Phys. Rev. Lett. 94, 237401 (2005).
[10] K. Kempa, Phys. Rev. B 66, 195406 (2002).
[11] S. M. Bose *et al.*, Phys. Rev. B 72, 153402 (2005).
[12] M. Lazzeri *et al.*, Phys. Rev. B 73, 155426 (2006).
[13] Z. Yao *et al.*, Phys. Rev. Lett. 84, 2941 (2000).
[14] J. Y. Park *et al.*, Nano Lett. 4, 517 (2004).
[15] E. Pop *et al.*, Phys. Rev. Lett. 95, 155505 (2005).
[16] M. Lazzeri *et al.*, Phys. Rev. Lett. 95, 236802 (2005).
[17] R. Krupke *et al.*, Science 301, 344 (2003).
[18] L. M. Huang *et al.*, J. Phys. Chem. B 108, 16451 (2004).
[19] M. Y. Sfeir *et al.*, Science 306, 1540 (2004).
[20] M. Y. Sfeir *et al.*, Science 312, 554 (2006).
[21] S. Reich *et al.*, *Carbon Nanotubes* (Wiley-VCH, 2004).
[22] S. M. Bachilo *et al.*, Science 298, 2361 (2002).
[23] C. Fantini *et al.*, Phys. Rev. Lett. 93, 147406 (2004).
[24] S. K. Doorn *et al.*, Appl. Phys. A 78, 1147 (2004).
[25] J. Maultzsch *et al.*, Phys. Rev. B 72, 205438 (2005).
[26] M. Y. Sfeir, Ph. D Thesis, Columbia University (2005).
[27] V. N. Popov, and L. Henrard, Phys. Rev. B 70, 115407 (2004).
[28] V. N. Popov *et al.*, Phys. Rev. B 72, 035436 (2005).
[29] S. Reich, and C. Thomsen, Phys. Rev. B 62, 4273 (2000).
[30] S. Reich *et al.*, Phys. Rev. B 64, 195416 (2001).
[31] M. Damnjanovic *et al.*, Phys. Rev. B 60, 2728 (1999).
[32] O. Dubay *et al.*, Phys. Rev. Lett. 88, 235506 (2002).
[33] S. Pisana *et al.*, Nature Mater. 6, 198 (2007).
[34] J. Yan *et al.*, Phys. Rev. Lett. 98, 166802 (2007).